\documentclass[12pt,a4paper,oneside]{article}
\usepackage[top=1in, bottom=1in, left=1.25in, right=1.25in]{geometry}
\usepackage{graphicx}
\usepackage{amsmath}
\usepackage{overcite}
\usepackage{url}
\usepackage[sort&compress,numbers,super]{natbib}

\title{\bf Semiclassical Dispersion Corrections efficiently improve Multi-Configurational Theory with Short-Range Density-Functional Dynamic Correlation}
\date{March 10, 2020}
\author{\vspace{0.3cm}Christopher J. Stein\footnotemark[2]
~and Markus Reiher\footnote{Corresponding author: markus.reiher@phys.chem.ethz.ch}\\
\textit{ETH Z\"urich, Laboratorium f\"ur Physikalische Chemie,}\\ 
\textit{Vladimir-Prelog-Weg 2, 8093 Z\"urich, Switzerland}\\
}

\begin{document}

\maketitle

\renewcommand*{\thefootnote}{\fnsymbol{footnote}}
\footnotetext{$\dagger$ present address: Department of Chemistry, University of California, Berkeley, CA 94720, USA}

\begin{abstract}
Multi-configurational wave functions are known to describe electronic structure across a Born-Oppenheimer surface qualitatively correct. 
However, for quantitative reaction energies, dynamical correlation originating from the many configurations involving excitations 
out of the restricted orbital space, the active space, must be considered.
Standard procedures involve approximations that eventually limit the ultimate accuracy achievable (most prominently, multi-reference
perturbation theory). At the same time, the computational cost increase dramatically due to the
necessity to obtain higher-order reduced density matrices. It is this disproportion that leads us here to propose
a MC-srDFT-D hybrid approach of semiclassical dispersion (D) corrections to cover long-range dynamical correlation
in a multi-configurational (MC) wave function theory which includes short-range (sr) dynamical correlation by
density functional theory (DFT) without double counting. We demonstrate that the reliability of this approach is very good (at negligible cost), 
especially when considering that standard second-order multi-reference perturbation theory usually overestimates dispersion interactions.
\end{abstract}

\newpage

\section{Introduction}

The accuracy required for most applications of electronic structure methods necessitates an efficient and reliable description of electron correlation.
Usually, electron correlation is divided into two classes: dynamic and non-dynamic (or static) electron correlation.
The former arises from the combined effect of the vast number of determinants (i.e., electron configurations) 
with small individual contribution to the energy, while the latter stems from the presence of several (near-)degenerate determinants
with large weights in the wave function expansion.

Although full configuration interaction (FCI) captures both types of correlation, its steep scaling with system size is prohibitive for almost all applications.
Various approaches have been developed to alleviate this problem and to provide approximate, but accurate models for
electronic structures dominated by static electron correlation.
Perturbation theory based selected CI methods,\cite{huro73,cimi87,mira93,nees03,tubm16,holm16,sharm17} ,
FCI quantum Monte Carlo (FCIQMC),\cite{boot09,boot10,shep12} and the density matrix renormalization group (DMRG)\cite{whit92,whit93,lege08,chan08,chan09,mart10,mart11,chan11,scho11,kura14,wout14,yana15,szal15,knec16,chan16}  
extended the range of application of static-correlation methods to large systems 
that were inaccessible for the traditional complete-active space self-consistent field (CASSCF) method.\cite{roos80,rued82,rued82a,rued82b,wern85,roos87,shep87}
Conceptually, selected CI methods aim at incorporating both static and dynamic correlation in a single calculation, whereas CAS-based methods such as CASSCF and DMRG aim for an accurate multi-configurational (MC) zeroth-order wave function that captures all static correlation 
with the dynamic correlation to be included in a separate step (e.g., by variants of
multi-configurational perturbation theory such as CASPT2\cite{ande90,ande92} or NEVPT2\cite{ange01,ange01a,ange02}). 

Multi-reference coupled-cluster variants\cite{jezi81,maha98,masi98,evan11,hana11,evan14,evan18} are another option when considering the fact 
that their single-configurational counterpart is a highly-accurate benchmark approach. However, the multitude of such variants already indicates
the conceptual intricacies that are associated with the generalization of coupled cluster theory toward a multi-reference situation.
More severe from the practical point of view is the increased computational effort. 
For example, the internally contracted multi-reference coupled-cluster with singles and doubles
suffers from the fact that reduced density matrices up to fifth order are required even with approximations\cite{hana11} and the number of terms in the amplitude equation is immense and cannot be efficiently reduced.
Very recently, a tailored coupled-cluster variant was proposed in combination with DMRG wave functions that reduces the multi-configurational coupled-cluster expressions to their single-configurational counterpart.\cite{veis16,veis18,faul19}
The multi-configurational effect is then included by constraining the amplitudes of excitations within the active space to those obtained from 
the DMRG calculation. Future work on this approach needs to show how universally applicable it actually is for 
strong multi-configurational cases that require large
active spaces and large orbital basis sets.

All of these approaches to include dynamical electron correlation contributions into the total electronic energy come at
a price which is the fact that they can easily exceed the computational effort to produce the CAS-type reference wave function. 
At the same time, the increased computational cost are not necessarily reflected in a significantly improved accuracy compared to that known from single-reference calculations.
For the calculation of dynamic correlation in single-configurational cases,
where the Hartree--Fock wave function is an appropriate zeroth-order approximation to the exact wave function and hence static correlation is absent,
highly accurate and efficient methods were developed.
For such explicitly correlated coupled cluster models, linear scaling with system size was achieved by exploiting the locality of electron correlation 
in molecules.\cite{schu01,tenn04,subo06,ziol10,roli13,ripl13,erik15,ma18}
By contrast, there is currently no method available for the accurate calculation of dynamical correlation for multi-configurational wave functions 
for large active orbital spaces that achieves the accuracy known for single-reference methods at moderate computational cost.

For multi-configurational calculations, the combination of CASSCF with CASPT2 or NEVPT2 is currently
the most common approach.
However, the necessity to calculate the four-particle density matrix elements
limits the size of the active space of the underlying CAS-type wave function to about 30 orbitals.
While this is a disadvantage in its own right, it is well-known that perturbation theory of second-order overestimates dispersion 
effects, especially for $\pi$--$\pi$ interactions.\cite{sinn02,cybu07,rile12,sedl13}
This is true for the single-configurational case and also holds for MC calculations (\textit{vide infra}).
In addition, M{\o}ller-Plesset perturbation theory of second order (MP2) is not accurate enough for many applications and it cannot 
be expected that its multi-reference analogs achieve a significantly higher accuracy.

In a situation where the accuracy and feasibility of explicitly correlated single-reference coupled cluster is unrivaled
by MC approaches and where quantitative MC methods achieve an accuracy that is comparable to MP2 (or slightly better) when applied to
single-reference cases, it is worthwhile considering approximate, but significantly more efficient schemes.
For this reason, MC-DFT hybrid theories that can efficiently pick up short-ranged (sr) dynamic correlations 
through srDFT\cite{stoll85,savi95,lein97,poll02,toul04,angy05,goll05,goll06,from07,hede15,hube16,hede18,fert19}
at reduced cost have been studied in recent years.
It is important to understand that the range separation in such MC-srDFT approaches affects the interelectronic distance in the two-electron integrals 
and avoids, by construction, double counting of correlation effects.\cite{stoll85,savi95,from07}
However, MC-srDFT still lacks long-range dynamical correlation, which is often dominated by correlations that produce attractive dispersion
interactions (note, however, that 'long-range' again refers to the interelectronic distance, whereas dispersion interactions are a concept
discussed for (usually large) internuclear distances and hence are an interpretation of the electron correlation effect). 

We note in passing that MC-pairDFT,\cite{lima14,gagl17}
which is a different combination of DFT and MC wave function theory not relying on (interelectronic) range separation but on a separate
treatment of Coulomb and exchange contributions, can to some extent recover dispersion interactions.\cite{bao19}
This is a somewhat surprising finding as such correlations are not expected to be contained in either the foundational CASSCF wave function 
(because of the restricted size of the active orbital space) or in the translated version of standard density functionals (because the
standard functionals suffer from the lack of such contributions to the correlation part of the functional).

In the domain of single-configurational calculations it has been demonstrated that semi-classical dispersion corrections with 
empirical parameters reliably describe dispersion at hardly any cost\cite{john05,grim04,grim06,grim10,grim11,cald17,cald19,sato09,sato10,tkat09,tkat12,stei10,stei11} compared
to the accuracy and computational cost associated with single-reference coupled cluster reference data.

In this letter, we argue that a CAS-type method that takes care of all static correlation effects combined with 
short-range dynamical electron correlation from DFT complemented by a long-range part modeled as semi-classical dispersion
corrections is a viable route toward accurate calculations for large systems with considerable multi-configurational character.
As an example, however, we choose to study seystems which {\it does not} feature any MC character at all so that short-range dynamical as well as static
correlation are suppressed and do not blur the effect of the long-range contribution. The distance-dependent interaction of two 
benzene molecules is such a case that isolates long-range correlation and is therefore our first system of choice.

\section{Dispersion and Dynamic Correlation in MC Wave Functions}
A CAS($N$,$L$)SCF wave function contains all determinants that are generated by distributing $N$ active electrons over $L$ (optimized) spatial
orbitals in all possible ways that conserve the spatial and spin symmetry of the wave function.
It is of practical importance, but also conceptually attractive, to keep this active space as compact as possible 
to have a clear separation of static and dynamic electron correlation.
The former is then understood to be fully embraced by the CAS-type wave function, whereas the contribution from the
latter is added in a subsequent calculation.
Apart from a few exceptions --- e.g. the double-shell effect of $3d$ transition metal complexes\cite{ande92a,pier01} or Rydberg states --- the active orbitals can be recruited from the valence-shell and are therefore already represented in a minimal basis set.\cite{stei19}

Dispersion interactions such as van der Waals forces arise from aligned polarizations being preferred over unaligned polarizations in molecular fragments.
They are a pure electron correlation effect that requires a substantial amount of virtual orbitals to be accurately accounted for.
The more compact the active space, the less dispersion interaction is included with the limit of complete absence of dispersion in the uncorrelated Hartree-Fock wave function.
A recent study\cite{gont17} revealed that a minimum of three carefully constructed virtual orbitals per occupied orbital is required to 
describe the majority of the two-body dispersion interactions.

The active orbitals of a CAS calculation, however, are selected and optimized to account for static correlation and do usually not include the fraction of the virtual space that would allow for a construction of such virtual orbitals that are optimal for the representation of dispersion.
There is therefore hardly any dispersion interaction included in a properly constructed CASSCF wave function 
and it must be corrected for in a separate calculation.

Two routes are, in principle, viable to accurately describe dispersion interaction for molecules with an MC wave function.
Almost exclusively, one of these routes is followed, i.e. choosing one method for the dynamic part of the electron correlation that describes dispersion interaction with sufficient accuracy such as multi-reference perturbation theory.
A second route, however, would be to make sure that dynamic correlation is calculated dispersion-free so that dispersion interactions 
can be added as an additional contribution to the total energy.

As noted above, semi-classical dispersion corrections proved to be exceptionally successful as an on-top correction for DFT calculations that also includes 
little to none dispersion interaction for most functionals.
Adding a similar correction to MC calculations is therefore a natural proposal if double counting of the dispersion interaction can be avoided.

In the case of a DMRG-optimized CAS-type wave function, which allows one to access active spaces of up to about 100 spatial orbitals,
it is, if in doubt, possible to eliminate any dispersion contribution to the energy between fragments from the wave function.
For this purpose, one needs to first optimize the orbitals in a converged DMRG-SCF calculation and subsequently localize the active orbitals. 
This can be realized with the Pipek-Mezey localization scheme\cite{pipe89} but any approach that ensures complete localization on only one fragment can be chosen.
In practical calculations of non-covalent interaction energies, however, already the optimized active orbitals are usually localized on a single fragment to a large extent.
After the localization, the wave function is optimized with DMRG-CI in this localized basis.
Since the dispersion contribution to the energy between two fragments $A$ and $B$ arises from a certain class of excited determinants\cite{thir15,schn16} (see Figure 1) in such a localized basis, the weights of these determinants can be determined from the DMRG wave function
after reconstructing the CI expansion coefficients.\cite{mori07,bogu11}
As a last step, a wave function is constructed that does not contain the dispersion determinants and the corresponding energy is evaluated.
The energy expectation value for this stripped wave function then provides a dispersion-free CASSCF type energy.
This approach can be classified as a dispersion free selected CI method.
While such a procedure is possible in principle, it can be assumed that the compactness of the active space avoids double-counting of dispersion effects in most cases so that the explicit correction will not be necessary.

\begin{figure}[h!]
\begin{center}
\includegraphics[width=0.6\textwidth]{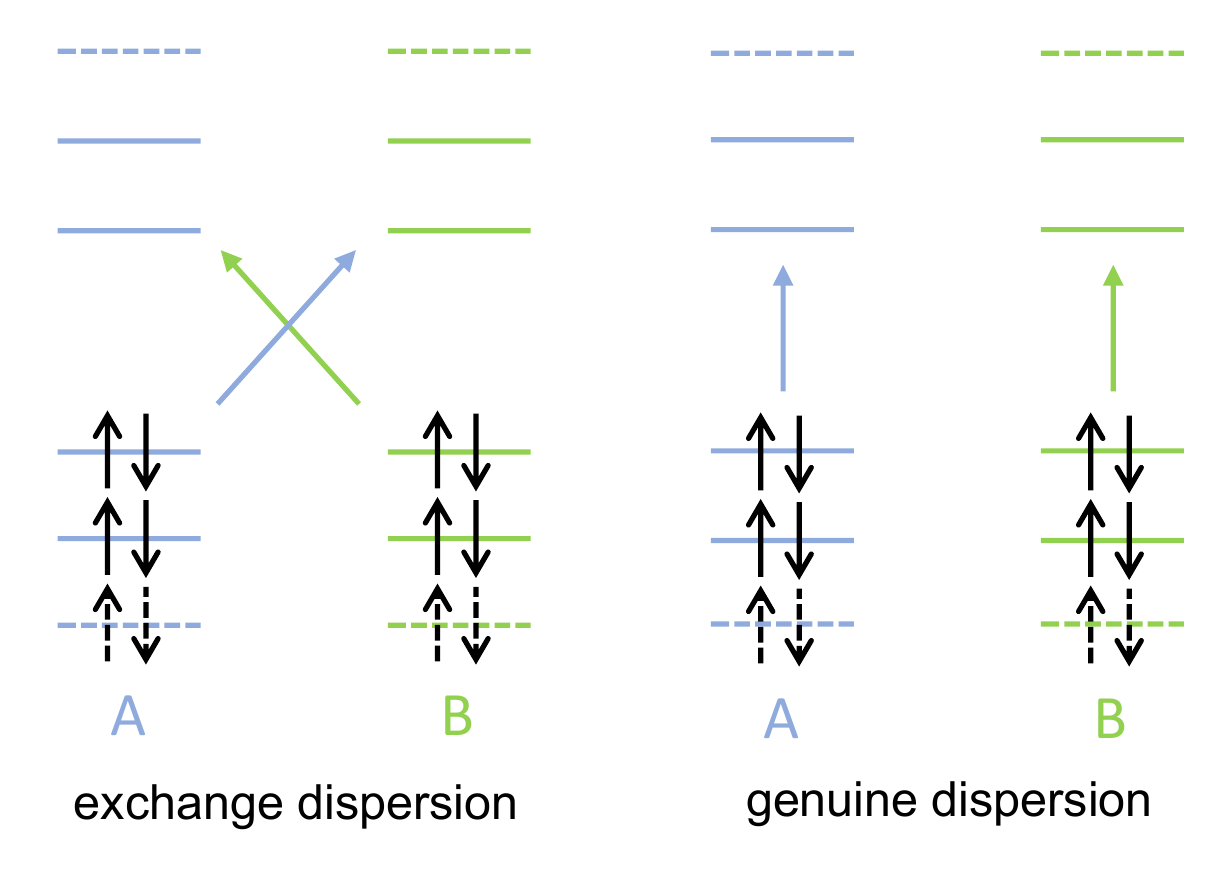}
\caption{Two classes of doubly excited determinants that define the dispersion interaction of the electron correlation. This assumes a perfect localization of the orbitals on the two interacting fragments $A$ and $B$.}
\label{fig:dispersion_benzene}
\end{center}
\end{figure}

Double counting can also be strictly avoided by range separation on the level of the two-electron integrals by means of a step function (typically the error function) and employing the short-range integrals for the srDFT calculation, whereas the long-ranged part is include in the CASSCF part (see references above).
Hence, dispersion effects are therefore completely absent in these calculations as neither CASSCF nor DFT incorporates them.
For this reason, it is both a straightforward and inexpensive approach to add semi-classical dispersion corrections to CASSCF/srDFT-type results
to account for the missing long-range electron correlation contribution, especially in view of
their tremendous success in the context of standard (single-determinant) Kohn-Sham DFT calculations.\cite{grim16} 

\section{Computational Details}

All multi-configurational calculations were carried out with a developer version of the {\sc Dalton} program.\cite{daltonpaper}
We employed a ANO-RCC-VTZ\cite{roos04,widm90} basis set in all calculations and the corresponding integrals were separated in a long-range and short-range part with a range-separation parameter of $\mu$~=~0.4~$a_0^{-1}$.
All srDFT calculations were carried out with the short-range version\cite{goll05,from07} of the PBE\cite{perd96} functional (srPBE).
We calculated the semi-classical D3 correction with the \textsc{DFT-D3} program\cite{dftd3} and Becke-Johnson damping\cite{grim11} and the standard PBE parameters as discussed below.
The proposed composite method is hence termed CASSCF-srPBE-D3.

All equilibrium structures and dissociation trajectories are part of the S22\cite{jure06} or S66 benchmark set\cite{reza11} for dispersion interactions.
The reference CCSD(T)/CBS calculations were also extracted from the original publications.

\section{Semiclassical Dispersion and the Benzene Dimer}

Considering the fact that (i) dispersion interactions are lacking in a CASSCF-srDFT setting due to a lack of long-range dynamic correlation, (ii) double-counting can be strictly avoided, and that (iii) such a model acquires efficiency through the assumption of a classical model, 
proceeding along these lines suggests the activation of semiclassical dispersion corrections.
The widely applied D3 dispersion correction with Becke-Johnson (BJ) damping,\cite{grim11}
\begin{align}
E^\mathrm{D3(BJ)} \sum_{I>J}\sum_{n=6,8} s_n \frac{C_{n,IJ}}{R^n_{IJ}+\left(a_1 \sqrt{C_{8,IJ}/C_{6,IJ}}+a_2 \right)^n} \, ,
\end{align}
is a suitable starting point,
where the first sum runs over all pairs of atoms $I$ and $J$, $R_{IJ}$ denotes the interatomic distance between these atoms, $C_{6,IJ}$ and $C_{8,IJ}$ are  functional-independent dispersion coefficients, the scaling factor $s_6$ is set to unity and $s_8$, $a_1$, and $a_2$ are the only functional-dependent parameters.
As we calculate the short-range part with a short-range variant of the PBE functional,\cite{perd96,goll05,from07}
we pick the D3 parameters for this functional for the evaluation of the dispersion energy contribution.

\begin{figure}[h!]
\begin{center}
\includegraphics[width=\textwidth]{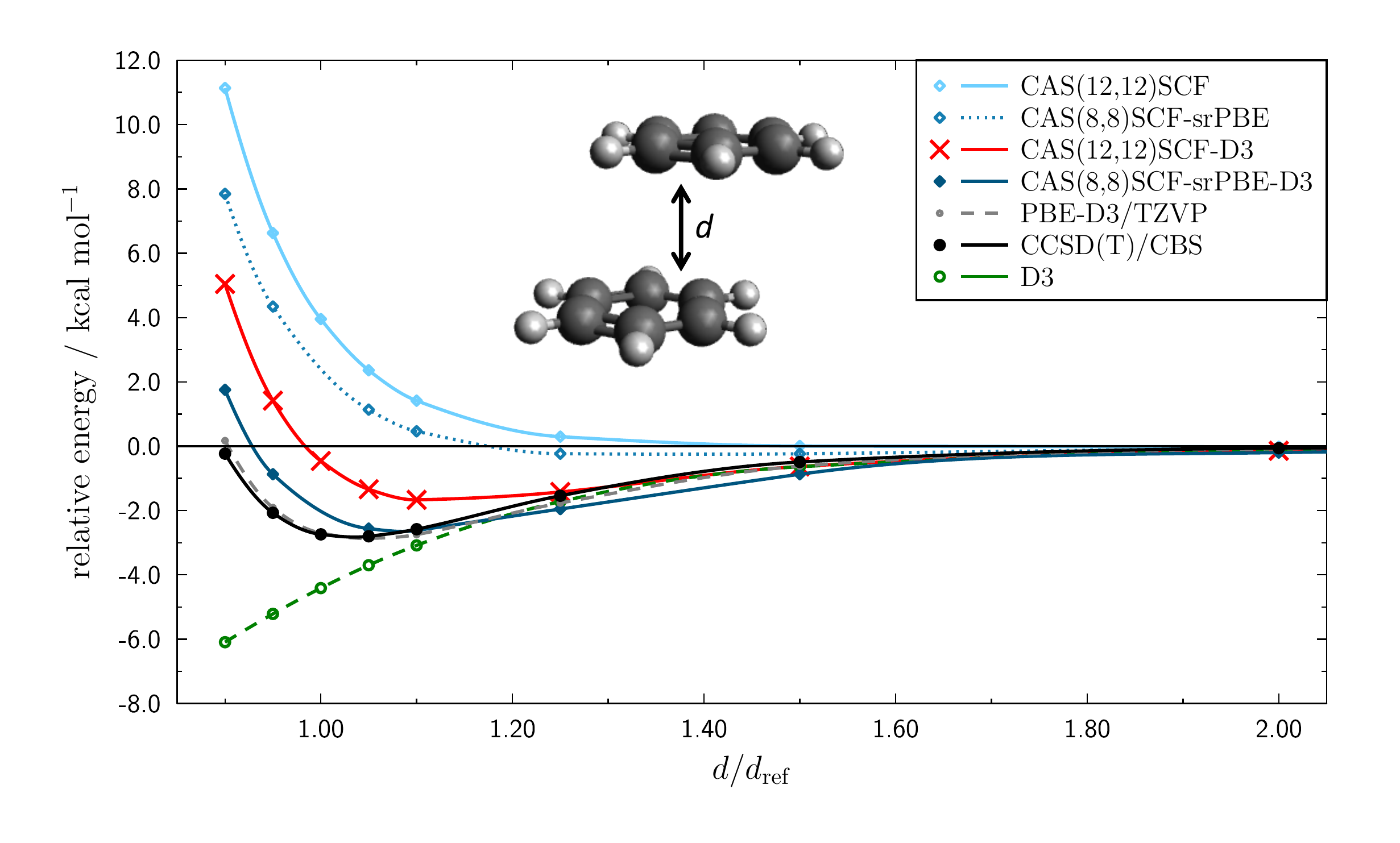}
\caption{Potential energy curves (in kcal/mol) of the benzene dimer in its stacked conformation calculated with different quantum chemical methods. 
The CCSD(T)/CBS curve serves as reference. The reference distance $d_\mathrm{ref}$\, is the shortest intermolecular distance obtained from the minimum structure of a MP2\cite{moll34,popl76} with a cc-pVTZ basis set.\cite{dunn89,kend92} The distance of the stacked carbon atoms at this equilibrium configuration is 3.403~\AA.}
\label{fig:dispersion_benzene}
\end{center}
\end{figure}

The systems we chose to demonstrate the capabilities of our composite approach needed to fulfill certain criteria.
The multi-configurational character must be large enough so that a stable active space can be identified, the bonding interaction must be purely or at least predominantly dispersion controlled in order to isolate dispersion interactions, and accurate benchmark data must be available.
The fraction of the S66 benchmark set\cite{reza11} that features strong $\pi$-$\pi$ interactions fullfills these criteria as the $\pi$-systems allow one to select stable active spaces and the bonding of these dimers is classified as predominantly dispersion mediated. 

Hence, the benzene dimer in a stacked conformation at varying inter-planar distances taken from the S22 benchmark set\cite{reza11} serves as an initial example for the analysis of our hybrid model.
We first calculate a CASSCF reference wave function with an active space that includes all twelve $\pi$-orbitals and electrons (CAS(12,12)SCF).
The corresponding potential energy curve (light blue line in Figure~\ref{fig:dispersion_benzene}) is repulsive as expected because the interaction is predominantly dispersive and not captured in this limited orbital space.
Next, we analyze the effect of dynamic correlation with a range-separated CASSCF-srPBE calculation.
We note that these calculations require only a smaller CAS(8,8) space because of a regularizing effect of the srDFT contribution.\cite{hede15}
This regularization leads to a decrease in entanglement \cite{lege03,riss06,bogu12,stei16}
for some of the orbitals such that they rotate out of the active space during the optimization and, hence, require us to exclude them from the active space.
The loss of entanglement is also reflected in the natural occupation numbers that change from 1.963 and 1.959 to 1.998 and 1.997 (two occupied orbitals with lowest occupation numbers) and from 0.036 and 0.036 to 0.002 and 0.002 (two unoccupied orbitals with largest occupation numbers) for a CASSCF calculation with and without srDFT, respectively, at the equilibrium structure.
The relative energies of these calculations are shown as the blue dotted line in Figure~\ref{fig:dispersion_benzene}. 
Although the repulsive part is shifted to shorter distances, no proper minimum is observed. 
The addition of the D3 correction to the CAS(12,12)SCF curve, however, shows a qualitatively correct picture with a minimum that is about half as deep as the correct one and slightly shifted to larger distances.
Finally, when combining CASSCF, srDFT, and the D3 correction (blue solid line in Figure~\ref{fig:dispersion_benzene}) the agreement with the reference results turns out to be very good, especially when considering that the D3 parameters were actually optimized for a PBE-only calculation.
In particular, the minimum energy agrees within 0.2 kcal/mol as opposed to the minimum of the CASPT2 curve (not shown here), where the minimum is almost twice as large.
Certainly, the agreement of D3 corrected PBE is even better (gray dashed line) but it must be emphasized that this molecule was part of the benchmark set for which the D3 parameters were optimized and there is also little MC character that could break the standard PBE model.
Moreover, if strong MC cases are to be considered, it can be anticipated that an MC-srDFT-D such as the CASSCF-srPBE-D3 model
studied here (possibly equipped with rigorous error bounds
from machine learning confidence intervals as described in Refs.\ \citenum{weym18,prop19}) will have 
a significant advantage over a standard single-configuration Kohn-Sham PBE-D3 calculation.

\section{Extended set of $\pi$-$\pi$ interacting systems}

\begin{table}[h!]
\caption{CASSCF-srPBE-D3 (comp) and CASSCF-D3 energies (in kcal/mol) for five different complexes calculated at the CCSD(T) optimized structure from the S66 benchmark set and for the interpolated CASSCF-srPBE-D3 minimum structure. The interplanar or plane/hydrogen distances $d$ are given in \AA \, and $\Delta$ denotes the difference to the reference values $E^x-E^\mathrm{CCSD(T)/CBS}$ in kcal/mol, where $x$ the denotes one of the dispersion corrected multiconfigurational methods.}
\label{table1}
\begin{tabular}{lrrrrrr}
\hline
&\multicolumn{5}{c}{\underline{CCSD(T)/CBS structure}}  \\
molecule &$d$& $E^\mathrm{CCSD(T)/CBS}$& $E^\mathrm{comp}$ & $\Delta$ & $E_\mathrm{diss}^\mathrm{CASSCF-D3}$& $\Delta$   \\
\hline
Pyridine dimer ($\pi$)       &3.70 & -3.90  & -3.00 & 0.90 & -1.15 & 2.75 \\
Benzene-Pyridine ($\pi$) &3.78 & -3.44  & -2.62 & 0.82 & -0.87 & 2.57  \\
Benzene dimer (T)           &2.55 & -2.88 & -2.56 & 0.32 & -1.48 & 1.40  \\
Pyridine dimer (T)            &2.59 & -3.54 & -3.02 & 0.52 & -1.95 & 1.59  \\
Benzene-Pyridine (T)       &2.52 & -3.33 & -2.99 & 0.34 & -1.85 & 1.48  \\
\hline
&\multicolumn{5}{c}{\underline{CASSCF-srPBE-D3 (comp) structure}} \\
molecule &$d$ & $E^\mathrm{comp}$ & $\Delta$ &&\\
\hline
Pyridine dimer ($\pi$)       & 3.98 & -3.52 & 0.38 \\
Benzene-Pyridine ($\pi$) & 4.07 & -3.13 & 0.31  \\
Benzene dimer (T)           & 2.72 & -2.79 & 0.08 \\
Pyridine dimer (T)            & 2.77 & -3.26 & 0.28 \\
Benzene-Pyridine (T)       &2.67  & -3.20 & 0.13 \\
\hline
\end{tabular}
\end{table}
We calculated the dimer dissociation curves for all combinations of benzene and pyridine in $\pi$-stacked or T-shaped configurations because their interaction energy, too, is classified as predominantly dispersion controlled, hence providing suitable test cases for our computational model.
As before, the active spaces were a CAS(12,12) including the whole $\pi$ space in case of the pure CASSCF calculations and a reduced CAS(8,8) for the CASSCF-srDFT calculations due to the regularization effect as discussed above.
The resulting dissociation energies are listed in Table~1.
Energy differences between the minimum structure of the reference calculations and the free monomers are labeled $E^\mathrm{CCSD(T)/CBS}$ and are evaluated for the CASSCF-srDFT-D3 composite approach (abbreviated as '\textit{comp}' in Table~\ref{table1}) and CASSCF-D3 models.
It is not surprising that in the pure CASSCF-D3 calculations, which lack short-range dynamical correlation, these energies are grossly underestimated in all cases.
We also note that the CASSCF-srPBE potential energy curves without dispersion correction (not listed in Table~\ref{table1}) are even unbound in most cases.

In addition, we calculated CASSCF-srDFT-D3 energies for three distances $d = a \cdot d_\mathrm{eq}$, with $a=1.05,1.1,1.25$ and with $d_\mathrm{eq}$ being the CCSD(T)/CBS equilibrium distance between the monomers, measured either as the interplanar distance ($\pi$-stacked dimers) or between the plane of the $\pi$-system and the H-atom that points toward the plane (T-shaped dimers).\cite{reza11} 
We fitted these energies to a Lennard-Jones type expression in order to approximately interpolate to the minimum of the CASSCF-srPBE-D3 curve.
The resulting minima are labeled 'CASSCF-srDFT-D3' in Table~\ref{table1} and --- as in the example of the $\pi$-stacked benzene dimer --- the intermolecular distances are larger than for the reference calculations.
The corresponding dissociation energies are still smaller than those calculated with the coupled cluster reference method.
These observations are more pronounced for the $\pi$-stacked complexes than for the T-shaped configurations.
In all cases, however, a significant stabilization of the dimer is correctly described when short-range dynamical correlation with srPBE and a long-range dispersion correction is added by means of the D3 model.
Remaining discrepancies with respect to the reference calculations can certainly be overcome by a refinement of the proposed model.

Overall, the potential energy curve for the benzene dimer calculated with CASSCF/srDFT including semiclassical D3 dispersion correction and the results for the other systems demonstrate that this approach has great potential to become an efficient multi-configurational model of sufficiently high accuracy, but negligible cost compared to the
CAS-type reference calculation and without double-counting of the different correlation contributions.
Naturally, there is plenty of room for improvement for subsequent work: 
i) Functional development --- it can be expected that the targeted development of short-range density functionals for the new model will further
improve its accuracy, 
ii) D3 parameters --- the parameters for the dispersion correction can be optimized for this method,
iii) D3 damping function adjustment as the damping function smoothly switches off the dispersion correction for small interatomic distances in order to remove any double counting with the exchange correlation functional, and
iv) optimization of new atomic orbital basis functions that are suitable for both srDFT and MC calculations.

\section{Conclusions}

The MC-srDFT-D hybrid model proposed here features 
i) short-range dynamical correlation described by DFT, 
ii) long-range static electron correlation described by a CAS-type wave function (e.g., optimized by DMRG, CASSCF, or FCIQMC),
and
iii) long-range dynamical correlation expressed in terms of semi-classical dispersion corrections.

No double-counting of the intermolecular dispersion energy is present in this model, because
the compactness of the active space from which the long-range MC wave function is constructed
prohibits any inclusion of dispersion interaction which is a dynamic electron correlation effect.
We note that already a simpler MC-D model that lacks the range separation step required for strictly avoiding double
counting (and therefore does not activate the srDFT part)
is likely to be an efficient and reliable model after a reparameterization of the semi-classical dispersion model in certain cases, 
too, given the fact that a properly chosen CAS will be compact and not plagued by
significant long-range dynamical correlations so that double counting is largely avoided.
We also note that active spaces are especially compact when they are selected with our recently developed automated active space selection protocol.\cite{stei16,stei16a,stei17,stei19}
In addition, any residual dispersion interaction can be eliminated from a CAS-type wave function with a correction scheme based on the exclusion 
of a certain subclass of determinants that is attributed to dispersion interaction as described above.

We should stress that there exist four options to improve on our model in a systematic way: development of more consistent short-range DFT functionals, reparameterization of the D3 parameters, an optimization of the damping scheme, and the development of new atomic orbital basis functions that are optimized for the MC-SrDFT model.
These improvements are intertwined and a parameter adjustment should involve all these options simultaneously.
An accuracy that rivals that of the standard CASSCF/CASPT2 or CASSCF/NEVPT2 approaches
is easily in reach and can most likely be exceeded with a properly parametrized 
MC-srDFT-D hybrid model, whereas the computational effort is negligible
once the CAS-type reference has been obtained.

\section{Acknowledgements}
We are grateful to Prof. Dr. H. J. Aa. Jensen for granting us access to an unpublished version of his short-range DFT implementation in the {\sc Dalton} program
that was has been part of the DMRG-srDFT approach.\cite{hede15}
Generous financial support by ETH Zurich is gratefully acknowledged.


\providecommand{\latin}[1]{#1}
\makeatletter
\providecommand{\doi}
  {\begingroup\let\do\@makeother\dospecials
  \catcode`\{=1 \catcode`\}=2 \doi@aux}
\providecommand{\doi@aux}[1]{\endgroup\texttt{#1}}
\makeatother
\providecommand*\mcitethebibliography{\thebibliography}
\csname @ifundefined\endcsname{endmcitethebibliography}
  {\let\endmcitethebibliography\endthebibliography}{}

\end{document}